\documentstyle[12pt]{article}
\begin{document}
\begin{titlepage}
\baselineskip=0.30in
\begin{flushright}
{\bf Revised version}

NUHEP-TH-96-3
\end{flushright}
\vspace{0.3in}

\begin{center}

{\Large One-loop QCD Corrections to Top Quark Decay into \\ 
a Neutralino and Light Stop}
\vspace{.5in}
 
     Chong Sheng Li{\footnote {On leave from Department of Physics, 
                           Peking University, China

		~~Electric address: csli@nuhep.phys.nwu.edu

	        ~~(After December: csli@svr.bimp.pku.edu.cn)}}, 
     Robert J. Oakes {\footnote{ Electric address: oakes@fnal.gov}}
and  Jin Min Yang{\footnote  {On leave from Department of Physics, 
                              Henan Normal University, China

                              ~~Electric address: jmyang@nuhep.phys.nwu.edu

                              ~~(After December:  yangjm@itp.ac.cn)}}

\vspace{.3in}
 
     Department of Physics and Astronomy, Northwestern University,\\
     Evanston, Illinois 60208

\end{center}
\vspace{.6in}

\begin{footnotesize}
\begin{center}\begin{minipage}{5in}

\begin{center} ABSTRACT\end{center}
 
We calculate the one-loop QCD corrections to $t\rightarrow \tilde t_1
\tilde \chi^0_j$ using dimensional reduction scheme, including QCD and 
supersymmetric QCD corrections. The analytic expressions for the corrections 
to the decay width are given, which can easily be extended to $t\rightarrow 
\tilde \chi^+_j \tilde b_i $. The numerical results show that the correction 
amounts to more than a 10\% reduction in the partial width relative to the 
tree level result. We also compare the corrections in the no-mixing stop case 
with those in the mixing stop case.
\end{minipage}\end{center}
\end{footnotesize}
\vspace{.8in}

PACS number: 14.80Dq; 12.38Bx; 14.80.Gt

\end{titlepage}
\eject
\baselineskip=0.28in

\begin{center}1. Introduction \end{center}

  The top quark has been dicovered by the CDF and D0 Collaborations at
the Fermilab Tevatron[1].  In the Standard Model
$t\rightarrow W^++b$ is the dominante decay mode. Beyond the SM, in addition to
the top decay into possible charged Higgs bosons plus bottom, a potentially
important decay channel of the top quark is the supersymmetric decay into a
lighter stop plus a neutralino, which has been extensively
discussed at tree level[2]. It is generally expected that the lighter
of the two stops is significantly lighter than the other squarks
because  the large top quark Yukawa coupling  drives the diagonal
stop masses to small values and enhances the off-diagonal mixing of
left-handed and right-handed stops, so the present squark mass
limits would not apply to the lighter stop. The best current lower bound on
the stop mass is  55GeV and comes from LEP, operating at $\sqrt s=130-140$ 
GeV[3]. The
D0 experiments at the Tevatron have excluded the existence of a
stop lighter than 100 GeV, albeit under certain assumptions[4]. 
Since the lightest
neutralino is the lightest supersymmetric particle the decay
$t\rightarrow \tilde t_1 \tilde \chi^0_1$ could occur in a reasonably large
 volume of the parameter space with a sizeable branching ratio[2].
The one-loop radiative corrections to both $t\rightarrow W^+b$ and
$t\rightarrow H^+b$ have been calculated [5,6] but the radiative
corrections to $t\rightarrow  \tilde t_1 \tilde \chi^0_j$
and $t\rightarrow \tilde \chi^+_j \tilde b_1$ have so far not been
calculated. In this paper we present the calculation of the
one-loop $O(\alpha_s)$ corrections to the
top quark decay into the lightest stop plus a neutralino, including both 
QCD and
supersymmetric QCD contributions. Our results can be generalized
straightforwardly to the decay $t\rightarrow \tilde \chi^+_j\tilde b_1$,
where $\tilde b_1$ is a light sbottom.
\vspace{1cm}

\begin{center}2. Tree-level \end{center}

In order to make this paper self-contained we first present 
the relevant
interaction Lagrangians of the Minimal Supersymmetric Standard Model
(MSSM) and the tree-level decay rates for $t\rightarrow \tilde t_1 \tilde \chi^0_j$. The
interactions of top and stop with neutralinos and gluinos are given by the
Lagrangians[7]
\begin{equation}
{\cal L}_{t\tilde t_i\tilde \chi^0_j}=-\sqrt 2 \bar t(L_{ij}P_L
+R_{ij}P_R) \tilde \chi^0_j \tilde t_i +h.c., 
\end{equation}
and
\begin{equation}
{\cal L}_{t\tilde t_i\tilde g}=-g_sT^a\bar t (a_i-b_i\gamma_5)\tilde g_a
\tilde t_i+h.c., 
\end{equation}
where
\begin{eqnarray}
a_1&=&\frac{1}{\sqrt 2} (\cos\theta-\sin\theta),
~~a_2=-\frac{1}{\sqrt 2} (\cos\theta+\sin\theta),\\
b_1&=&-\frac{1}{\sqrt 2} (\cos\theta+\sin\theta),
~~b_2=\frac{1}{\sqrt 2} (\sin\theta-\cos\theta),\\
L_{1j}&=&A_j\cos\theta-C_j\sin\theta,
~~L_{2j}=-A_j\sin\theta-C_j\cos\theta,\\
R_{1j}&=&B_j\cos\theta-D_j\sin\theta,
~~R_{2j}=-B_j\sin\theta-D_j\cos\theta
\end{eqnarray}
with 
\begin{eqnarray}
A_j&=&D_j^*=\frac{gm_tN^*_{j4}}{2m_W\sin\beta},
~~B_j=C_j^*+\frac{gN'_{j2}}{2C_W},\\
C_j&=&\frac{2}{3}eN'^*_{j1}-\frac{2}{3}\frac{gS_W^2}{C_W}N'^*_{j2},
\end{eqnarray}
and
\begin{equation}
N'_{j1}=N_{j1}C_W+N_{j2}S_W,
~~N'_{j2}=-N_{j1}S_W+N_{j2}C_W,
\end{equation}
Here $S_W\equiv\sin\theta_W, C_W\equiv\cos\theta_W$,
$P_{L,R}\equiv\frac{1}{2}(1\mp\gamma_5)$,
and $ N_{ij}$ are the elements of the $4\times4$ matrix $N$ defined 
in Ref.[7], which can be calculated numerically. 
$T^a=\lambda^a/2$ are the Gell-Mann matrices and  $\theta$ is the mixing angle
between left- and right-handed stops
which are related to the mass eigenstates $\tilde t_i$
in Eqs. (1) and (2) by
\begin{eqnarray}
\left ( \begin{array}{l} \tilde t_1\\ \tilde t_2 \end{array} \right )
=\left ( \begin{array}{ll} \cos\theta & \sin\theta\\
           -\sin\theta & \cos\theta \end{array} \right )
\left ( \begin{array}{ll} \tilde t_L\\ \tilde t_R \end{array} \right ).
\end{eqnarray}
This rotation matrix, Eq. (10), diagonalizes the stop mass matrix[8]
\begin{eqnarray}
M^2_{\tilde t}=\left ( \begin{array}{cc}
M^2_{\tilde t_L}+m_t^2+0.35\cos(2\beta)M_Z^2 &-m_t (A_t+\mu\cot\beta)\\
-m_t (A_t+\mu\cot\beta) & M^2_{\tilde t_R}+m_t^2+0.16\cos(2\beta)M_Z^2
\end{array} \right ),
\end{eqnarray}
where $ M^2_{\tilde t_L}, M^2_{\tilde t_R}$ are the soft SUSY-breaking mass
terms for left- and right-handed stops,
$\mu$ is the supersymmetric Higgs mass parameter in the superpotential, 
$A_t$ is the trilinear soft SUSY-breaking parameter,
and $\tan\beta=v_2/v_1$ is the ratio of the
vacuum expectation values of the two Higgs doublets.
              
The tree-level Feynman diagram for the decay 
$t\rightarrow \tilde t_1 \tilde \chi_j^0$ is shown in Fig.1(a), and the
tree-level partial decay width is given by
\begin{eqnarray}
\Gamma_0&=&\frac{1}{16\pi m_t^3}
\lambda^{1/2}(m_t^2,m^2_{\tilde \chi^0_j},m^2_{\tilde t_1})\left [
(\vert L_{1j}\vert^2+\vert R_{1j}\vert^2)
(m_t^2+m^2_{\tilde \chi^0_j}-m^2_{\tilde t_1})\right.\nonumber\\
& & \left.+4Re(L_{1j}^*R_{1j})m_tm_{\tilde \chi^0_j}\right ]
\end{eqnarray}
where $\lambda(x, y, z)=(x-y-z)^2-4yz$.
\vspace{1cm}

\begin{center}3. Virtual corrections \end{center}

Since the conventional dimensional regularization violates supersymmetry,
in our calculation we will use dimensional reduction technique[9], 
which preserves 
supersymmetry, for regularization of the ultraviolet divergences in the 
virtual loop corrections, although there is only a small difference between 
the both schemes to first order in the QCD and weak couplings.
In fact, in dimensional reduction scheme, at the one-loop level the 
$\epsilon$-scalars convert the dimensional
regularization result to the result which would be obtained by simply performing
the numerator algebra in four dimensions[9].
To regulate the infrared divergences associated with soft and collinear gluon
emission we will give the gluon a small finite mass $\lambda$ which is 
legitimate for our purposes
since the non-Abelian nature of QCD does not show up in  this order.
We will also adopt the on-shell renormalization scheme[10] in which the coupling
constant and the physical masses are chosen to be the renormalized
parameters. The finite parts of the counterterms are then fixed by the
renormalization conditions that the quark and the squark propagators have
poles at their physical masses. For the QCD and SUSY-QCD corrections to the
decay $t\rightarrow \tilde t_1\tilde \chi ^0_j$, which we are considering,
only the top quark mass and the stop mixing angle in the bare
coupling need to be renormalized. By introducing appropriate counterterms
the renormalized amplitude can be expressed as
\begin{equation}
M_{ren}=-i\sqrt {2}\bar u(\tilde \chi^0_j)(aP_R+bP_L)u(t) 
\end{equation}
with
\begin{eqnarray}
a&=&L^*_{1j}+\delta L^*_{1j}
+L^*_{1j}(\frac{1}{2}\delta Z^R_t+\frac{1}{2}\delta Z_{11})
+L^*_{2j}\delta Z_{12}
+\Lambda^{QCD}_R+\Lambda^{SUSY-QCD}_R,\\
b&=&R^*_{1j}+\delta R^*_{1j}
+R^*_{1j}(\frac{1}{2}\delta Z^L_t+\frac{1}{2}\delta Z_{11})
+R^*_{2j}\delta Z_{12}+\Lambda^{QCD}_L+\Lambda^{SUSY-QCD}_L,
\end{eqnarray}                                                           
where $\Lambda^{QCD}_{L,R}$ and $\Lambda^{SUSY-QCD}_{L,R}$ are the vertex
corrections from the irreducible vertex diagrams,
expressions for which will be given below.
 $\delta L^*_{1j}$ and $\delta R^*_{1j}$ are the shifts from the
bare couplings to renormalized couplings and , as mentioned above, can be
found by renormalizing the top quark mass and the stop mixing angle:
\begin{eqnarray}
\delta L^*_{1j}&=&L^*_{2j}\delta \theta + L^{*(m_t)}_{1j}
                 \frac{\delta m_t}{m_t},\\
\delta R^*_{1j}&=&R^*_{2j}\delta \theta + R^{*(m_t)}_{1j}\frac{\delta m_t}{m_t},\\
L^{*(m_t)}_{1j}&=&A^*_j\cos\theta, ~~R^{*(m_t)}_{1j}=-D^*_j\sin\theta.
\end{eqnarray}
The counterterms and the renormalization constants in Eqs.(14)-(17)
are defined by
\begin{eqnarray}
m_t^0&=&m_t+\delta m_t, \\
\theta ^0&=&\theta + \delta \theta,\\
t^0&=&Z_t^{1/2}t=(1+\delta Z^L_tP_L + \delta Z^R_tP_R)^{1/2}t,
\end{eqnarray}
and 
\begin{equation}
\tilde t_1^0=(1+\delta Z_{11})^{1/2} \tilde t_1+\delta Z_{12}\tilde t_2.
\end{equation}
Calculating the self-energy diagrams for the top quark in Figs. 1(b) and 2(a)
we obtain
\begin{eqnarray}
\frac{\delta m_t}{m_t}&=&\frac{\alpha _s C_F}{4\pi}
\left[-2\Delta+4F_0^{(ttg)}-2F_1^{(ttg)}
-\frac{m_{\tilde g}}{m_t}\alpha_{ii}F_0^{(t\tilde g \tilde t_i)}
-\sigma_{ii}F_1^{(t\tilde g \tilde t_i)}\right ],\\ 
\delta Z_t^L&=&\frac{\alpha _s C_F}{4\pi}
\left[-2\Delta+2F_1^{(ttg)}+m_t^2(4G_1^{(ttg)}-8G_0^{(ttg)})
+(\sigma_{ii}-\lambda_{ii})F_1^{(t\tilde g \tilde t_i)}\right.\nonumber\\
& & \left.+2m_t^2\sigma_{ii}G_1^{(t\tilde g \tilde t_i)}
+2m_tm_{\tilde g}\alpha_{ii}G_0^{(t\tilde g \tilde t_i)}
\right ], 
\end{eqnarray}
and
\begin{eqnarray}
\delta Z_t^R&=&\frac{\alpha _s C_F}{4\pi}
\left[-2\Delta+2F_1^{(ttg)}+m_t^2(4G_1^{(ttg)}-8G_0^{(ttg)})
+(\sigma_{ii}+\lambda_{ii})F_1^{(t\tilde g \tilde t_i)}\right.\nonumber\\
& &\left. +2m_t^2\sigma_{ii}G_1^{(t\tilde g \tilde t_i)}
+2m_tm_{\tilde g}\alpha_{ii}G_0^{(t\tilde g \tilde t_i)}
\right ],
\end{eqnarray}
where the sum over $i(=1,2)$ is implied, and 
\begin{eqnarray}
\sigma_{ij}&=&a_ia_j+b_ib_j,\\
\alpha_{ij}&=&a_ia_j-b_ib_j,\\
 \lambda_{ij}&=&a_ib_j+a_jb_i,\\  
F^{(ijk)}_n&=&\int^1_0 dy y^n\log \left [\frac{m_i^2y(y-1)+m^2_j(1-y)
+m^2_k y}{\mu ^2}\right ], 
\end{eqnarray}
and
\begin{equation}
G^{(ijk)}_n= -\int^1_0 dy \frac{y^{n+1}(1-y)}{m_i^2y(y-1)+
m^2_j(1-y)+m^2_ky}.
\end{equation}
Here, $\Delta\equiv \frac{1}{\epsilon}-\gamma_E+\log 4\pi$ 
with $\gamma_E$ being the Euler constant and $D=4-2\epsilon$ is
the space-time dimension. The color
factor $C_F=4/3$ for $SU(3)$ and $\mu$ is the 't Hooft mass parameter in the
dimensional regularization scheme.
Similarly, from Fig. 1(c), 2(b) and 2(d), one finds, for the stop,
\begin{eqnarray}
\delta Z_{11}&=&\frac{\alpha_s C_F}{4\pi}\left [ 
-F_0^{(\tilde t_1 \tilde t_1 g)}-2F_1^{(\tilde t_1 \tilde t_1 g)}
-2m^2_{\tilde t_1}
(G_0^{(\tilde t_1 \tilde t_1 g)}+G_1^{(\tilde t_1 \tilde t_1 g)})\right.
			\nonumber\\
& & \left. +4[F_1^{(\tilde t_1  t \tilde g)}
+m^2_{\tilde t_1}G_1^{(\tilde t_1  t \tilde g)}
-m_t^2G_0^{(\tilde t_1  t \tilde g)}
+m_tm_{\tilde g}\sin(2\theta)G_0^{(\tilde t_1  t \tilde g)} \right ],
\end{eqnarray}
and
\begin{eqnarray}
\delta Z_{12}&=&\frac{\alpha_s C_F}{4\pi}\cos(2\theta)\left \{ \left (
\frac{4m_tm_{\tilde g}}{m^2_{\tilde t_2}-m^2_{\tilde t_1}}
+\sin(2\theta)\right ) \Delta \right.\nonumber\\
& & \left.+\frac{1}{m^2_{\tilde t_1}-m^2_{\tilde t_2}}\left [
\sin(2\theta) (\bar A_0(m_{\tilde t_1})-\bar A_0(m_{\tilde t_2}))
+4m_tm_{\tilde g}F_0^{(\tilde t_1  t \tilde g)}\right ] \right \}
\end{eqnarray}
with  
\begin{equation}
\bar A_0(m)=m^2\left [1-\log (\frac{m^2}{\mu^2}) \right ].
\end{equation}

We have fixed the wave function renormalization constants and the top quark
mass counterterm by the on mass-shell renormalization scheme condition.
The mixing angle counterterm is fixed by the requirement that $\delta\theta$
 exactly cancel the remainder of the sum of all the ultraviolet(UV) divergent 
terms in the square of the renormalized ampltude, insuring the UV finiteness
of the physical observables. 
From this requirement we found  that the 
mixing angle counterterm simply is the negative 
of the counterterm $\delta Z_{12}$; that is, 
\begin{equation}
\delta \theta=-\delta Z_{12}.
\end{equation}
This condition insures that all the ultraviolet divergences will cancel in the virtual corrections to
the decay width, as will be seen below, and 
 is in agreement with Ref.[11].

  The calculations of the irreducible vertex corrections 
from Fig. 1(d) and  2(c) results in
\begin{eqnarray}
\Lambda ^{QCD}&=&\Lambda^{QCD}_LP_L+\Lambda^{QCD}_RP_R\nonumber\\
&=&\frac{\alpha_s C_F}{4\pi}\left \{ (L^*_{1j}P_R
+R^*_{1j}P_L)[\Delta+ 4\bar C_{24} \right.\nonumber\\
& & +m^2_t(2C_0+2C_{11}-C_{12}+C_{21}-C_{23})
+m^2_{\tilde t_1}(2C_0+2C_{11}\nonumber\\
& &  +C_{12}+C_{23})-m^2_{\tilde \chi ^0_j}(2C_0+
2C_{11}-C_{12}-C_{22}+C_{23})]\nonumber\\
& & \left.+2(L^*_{1j}P_L+R^*_{1j}P_R)m_t
m_{\tilde \chi^0_j}(C_{11}-C_{12})\right \}
(-p, k_1,\lambda, m_t, m_{\tilde t_1}),
\end{eqnarray}
and
\begin{eqnarray}
\Lambda ^{SUSY-QCD}&=&\Lambda ^{SUSY-QCD}_LP_L+\Lambda ^{SUSY-QCD}_RP_R
                    \nonumber\\
&=&\frac{\alpha_s C_F}{4\pi}\left \{  [(L_{2j}-L_{2j}
\cos 2\theta-L_{1j}\sin 2\theta)P_L  \right. \nonumber\\
& & +(-R_{2j}-R_{2j}\cos 2\theta-R_{1j}\sin
2\theta)P_R ]\Delta \nonumber\\
& & +\left ( S^{(1)}_{ji} [4\bar C_{24}+m_t^2(C_{21}-C_{23}
+C_{11}-C_{12})\right.  \nonumber\\
& &  +m^2_{\tilde t_1}(C_{22}-C_{23})+m^2_{\tilde \chi ^0_j}(
C_{23}+C_{12}) ]  \nonumber\\
& & +m_t^2S^{(4)}_{ji}(C_{11}-C_{12}+C_0)+m_tm_{\tilde \chi^0_j}
 [S^{(3)}_{ji}C_{12}+S^{(2)}_{ji}(C_0+C_{11}) ] \nonumber\\
& &\left. +m_{\tilde g}m_{\tilde \chi^0_j}
S^{(7)}_{ji}(C_0+C_{12})+m_tm_{\tilde g}[S^{(6)}_{ji}C_0+S^{(8)}_{ji}(C_{11}
-C_{12})]\right )P_R \nonumber\\
& & +\left (S^{(2)}_{ji}[4\bar C_{24}
    +m_t^2(C_{21}-C_{23}+C_{11}-C_{12})
    +m^2_{\tilde t_1}(C_{22}-C_{23}) \right.\nonumber\\
& & +m^2_{\tilde \chi ^0_j}(C_{23}+C_{12})]
    +m_t^2S^{(3)}_{ji}(C_{11}-C_{12}+C_0)
    +m_tm_{\tilde \chi ^0_j}[S^{(4)}_{ji}C_{12} \nonumber\\
& & +S^{(1)}_{ji}(C_0+C_{11})+m_{\tilde g}m_{\tilde \chi ^0_j}S^{(8)}_{ji}
    (C_0+C_{12}) \nonumber\\
& & \left.\left. +m_tm_{\tilde g}[S^{(5)}_{ji}C_0+S^{(7)}_{ji}(C_{11}-C_{12})]
\right ) P_L\right \}(-p, k_2,m_{\tilde t_i},m_{\tilde g}, m_t),
\end{eqnarray}
respectively, where  the sum over $i(=1,2)$ is implied. In Eqs.(35) and (36)
\begin{eqnarray}
S^{(1)}_{ji}&=&(\alpha_{1i}+\beta_{1i})R_{ij}, 
~~S^{(2)}_{ji}=(\alpha_{1i}-\beta_{1i})L_{ij}, \nonumber\\
S^{(3)}_{ji}&=&(\alpha_{1i}+\beta_{1i})L_{ij}, 
~~S^{(4)}_{ji}=(\alpha_{1i}-\beta_{1i})R_{ij}, \nonumber\\
S^{(5)}_{ji}&=&(\sigma_{1i}-\lambda_{1i})L_{ij}, 
~~S^{(6)}_{ji}=(\sigma_{1i}+\lambda_{1i})R_{ij} \nonumber\\
S^{(7)}_{ij}&=&(\sigma_{1i}+\lambda_{1i})L_{ij}, 
~~S^{(8)}_{ji}=(\sigma_{1i}-\lambda_{1i})R_{ij}, 
\end{eqnarray}
where
$\beta_{ij}=a_ib_j-b_ia_j$, and $C_0$, $C_{ij}$
are the three-point Feynman integrals given in the appendices of Ref. [12].

The virtual correction to the decay rate is then
\begin{eqnarray}
\delta \Gamma_{virt}&=&\frac{1}{16\pi m_t^3}
\lambda^{1/2}(m_t^2,m^2_{\tilde \chi^0_j},m^2_{\tilde t_1})
Re \left \{ 2(m_t^2+m^2_{\tilde \chi^0_j}-m^2_{\tilde t_1})
\left [( L_{1j} L_{2j}^*+R_{1j} R_{2j}^*)(\delta \theta+
\delta Z_{12})\right.\right. \nonumber\\
& & +( L_{1j} L_{1j}^{*(m_t)}+R_{1j} R_{1j}^{*(m_t)})\frac{\delta m_t}{m_t}
+(\vert L_{1j}\vert^2+\vert R_{1j}\vert^2)(\frac{1}{2}\delta Z_{11}
 +\delta_0^{QCD}) \nonumber\\
& & \left. +\frac{1}{2}(\vert L_{1j}\vert^2\delta Z_t^R
+\vert R_{1j}\vert^2\delta Z_t^L)
+( L_{1j}S^{(1)}_j+ R_{1j}S^{(2)}_j)\delta_0^{SUSY-QCD} 
+ L_{1j}\delta_1+R_{1j}\delta_2 \right ]  \nonumber\\
& & +4m_tm_{\tilde \chi^0_j}\left [( L_{1j} R_{2j}^*+R_{1j} L_{2j}^*)
 (\delta \theta+\delta Z_{12})
 +( L_{1j} R_{1j}^{*(m_t)}+R_{1j} L_{1j}^{*(m_t)})\frac{\delta m_t}{m_t}
     \right. \nonumber\\
& &
+(L_{1j} R_{1j}^*+R_{1j} L_{1j}^*)(\frac{1}{2}\delta Z_{11}+\delta_0^{QCD})
+\frac{1}{2}(L_{1j} R_{1j}^*\delta Z_t^L
  +R_{1j} L_{1j}^*\delta Z_t^R) \nonumber\\
& & 
\left. \left.
+( L_{1j}S^{(2)}_j+ R_{1j}S^{(1)}_j)\delta_0^{SUSY-QCD} 
+ L_{1j}\delta_2+R_{1j}\delta_1 \right ]\right \} ,
\end{eqnarray}
where $\delta_0^{QCD}$ and $\delta_0^{SUSY-QCD}$ are the UV divergent parts
 of the QCD
and SUSY-QCD vertex corrections, respectively.
These are given by
\begin{equation}
\delta_0^{QCD}=\delta_0^{SUSY-QCD}=\frac{\alpha_s C_F}{4\pi} \Delta,
\end{equation}
and $\delta_1, \delta_1, S^{(1)}_j$ and $S^{(2)}_j$ are  defined to be
\begin{eqnarray}
\delta_1&=&(\Lambda_R^{QCD}+\Lambda_R^{SUSY-QCD})_{\rm finite},\\
\delta_2&=&(\Lambda_L^{QCD}+\Lambda_L^{SUSY-QCD})_{\rm finite},\\
S^{(1)}_j&=&-R_{2j}-R_{2j}\cos 2\theta-R_{1j}\sin 2\theta,
\end{eqnarray}
and
\begin{eqnarray}
S^{(2)}_j&=&L_{2j}-L_{2j}\cos 2\theta-L_{1j}\sin 2\theta.
\end{eqnarray}
We have checked analytically that all the ultraviolet divergences  
indeed cancel in the virtual corrections to
the decay width,  but the infrared divergent terms  presist.

\vspace{1cm}

\begin{center}4. Real corrections \end{center}

As is well known[13], to cancel the infrared divergences 
in the virtual corrections one needs to include real gluon emission,
namely,  $t\rightarrow \tilde t_1 \tilde \chi^0_j g$, as shown in Figs.1(e,f).
As above, we will regulate the infrared divergences associated with the soft and
collinear real gluon emission by the same finite small gluon mass $\lambda$.
In the calculation of the  corrections due to real gluon emission 
to the partial width, it was necessary to 
perform the integration over the three-body phase space. 
 After tedious but straightfoward calculations we obtained
\begin{eqnarray}
\delta \Gamma_{real}&=&\frac{\alpha_s C_F}{4\pi}\frac{1}{2\pi m_t}\left \{
(\vert L_{1j}\vert ^2+\vert R_{1j}\vert ^2) 
 [I+I^1_0-2(m_t^4-(m^2_{\tilde \chi^0_j}-m^2_{\tilde t_1})^2)I_{01}\right.
 \nonumber\\
& & +2(m^2_{\tilde t_1}-m_t^2-m^2_{\tilde \chi^0_j})(I_0+I_1+m_t^2I_{00}
+m^2_{\tilde t_1}I_{11})] \nonumber\\
& & \left. +8m_t m_{\tilde \chi^0_j} Re(L^*_{1j}R_{1j}) 
[(m^2_{\tilde \chi^0_j}-m_t^2-m^2_{\tilde t_1})I_{01}
-m^2_{\tilde t_1}I_{11}-m_t^2 I_{00}-I_0-I_1]\right \},
\end{eqnarray}
Here we adopt the
notation of Ref.[14] where the definition of the functions 
$I_i, I_{ij}(m_t, m_{\tilde t_1},m_{\tilde \chi^0_j})$ can be found.
We also have checked numerically that the infrared divergences  in
$\delta \Gamma_{real}$ and $\delta \Gamma_{virt}$ do indeed cancel.   
\vspace{1cm}

\begin{center}5. Numerical results and discussions\end{center}

In the following we give the numerical results for 
$t\rightarrow \tilde t_1 \tilde \chi^0_1$, where $\tilde \chi^0_1$ is the 
lightest neutralino.  In our numerical calculation we fixed $M=200$ GeV,
$\mu=-100$ GeV and we used the relation 
$M^{\prime}=\frac{5}{3}\frac{g'^2}{g^2} M$ [7]
to fix $M'$. For the parameters in stop sector 
 we assumed $M_{\tilde t_R}=M_{\tilde t_L}$ and took the combination
 $A_t+\mu\cot\beta$ to be one parameter. Note that $(A_t+\mu\cot\beta)=0$
corresponds to the case of no mixing in the stop mass matrix, Eq.(11).
There are then three free parameters in the stop sector 
and  we chose $m_{\tilde t_1}, \tan\beta$ , and  $(A_t+\mu\cot\beta)$ as the
three independent parameters.     
Other input parameters are $m_Z=91.188 GeV,  ~\alpha_{em}=1/128.8 $,
and $G_F=1.166372\times 10^{-5}(GeV)^{-2}$. The $W$ mass was determined from
$[15]$
\begin{equation}
m_W^2(1-\frac{m_W^2}{M_Z^2})=\frac{\pi\alpha}{\sqrt 2 G_F}\frac{1}{1-\Delta r},
\end{equation}
where, for a heavy top,  $\Delta r$ is given by [16]
\begin{equation}
\Delta r\sim -\frac{\alpha N_Cc_W^2m_t^2}{16\pi^2s_W^4m_W^2}.
\end{equation}

Figure 3 shows the relative correction  to 
the decay rate $\delta \Gamma/\Gamma_0 $, $\Gamma_0 $ being the tree-level rate,
as a function of the  
lighter stop mass assuming $m_{\tilde g}=500$GeV and $\tan\beta=11$. 
The solid curve in Fig.3  corresponds to  $A_t+\mu\cot\beta=0$, 
 the no-mixing case, 
while the dotted curve corresponds to $A_t+\mu\cot\beta=100$GeV, a
mixing case. Note that in Fig.3 the lightest neutralino mass 
is $m_{\tilde \chi^0_1}=68$ GeV. It is clear  that
the correction in the mixing case is larger than in the no-mixing case 
and can reach -20\% for $m_{\tilde t_1}=100$ GeV. 
Figure 4 shows the dependence of the relative correction to the decay width
 on the value of gluino mass for $m_{\tilde t_1}=50$ GeV. Other parameter
values are the same as in Fig.3. For the solid curve $m_{\tilde t_1}=50$ GeV
and $m_{\tilde t_2}=64$ GeV and there are  two peaks at 
$m_{\tilde g}=112$ GeV and  $m_{\tilde g}=126$ GeV
 due to the fact that  $m_t=176$ GeV
 and the threshod for open top decay into 
gluino and  stop is crossed in these regions. 
For the dashed curve $m_{\tilde t_1}=50$ GeV
and $m_{\tilde t_2}=194$ GeV  and  there is  only one peak at 
$m_{\tilde g}=126$ GeV.
When the gluino mass
is heavier than 200 GeV the correction in the mixing case is larger
than in the no-mixing case and both corrections increase with gluino mass.  
Decoupling effects do not occur here, in contrast to the virtual
SUSY corections to the decay and production processes in the SM. 
In Figure 5 we present the dependence of the relative correction to the decay 
width on the value of $\tan\beta$ assuming $m_{\tilde g}=500$GeV, 
$m_{\tilde t_1}=50$ GeV and $A_t+\mu\cot\beta=100$ GeV.
Only in the region where $\tan\beta<2$ is the  correction to the decay width
very sensitive to  the value of $\tan\beta$.    

In conclusion, we have shown that the one-loop QCD and SUSY-QCD corrections to 
$t\rightarrow \tilde t_1\tilde \chi^0_j$ can exceed -10\% of the tree level 
partial width in both the no-mixing and the  mixing case of stop masses, 
and these corrections are not sensitive to $\tan\beta$ for $\tan\beta>2$.
\vspace{.5cm}

Note added: While preparing this manuscript the preprint of A.Djouadi, 
W.Hollik and C.Junger (hep-ph/9605340) appeared where the QCD correction 
to the process $t\rightarrow \tilde t_1 \tilde \chi^0_j$ is also calculated.
But Eq.(14) of their original paper were not correct. Very recently,
in their revised version
this mistake has been corrected by them. We thank W.Majerotto for useful 
communication.  
\vspace{.5cm}

This work was supported in part by the U.S. Department of Energy, Division
of High Energy Physics, under Grant No. DE-FG02-91-ER4086. 
\eject

{\LARGE References}
\vspace{0.3in}
\begin{itemize}
\begin{description}
\item[{\rm [1]}] CDF Collaboration,  Phys.Rev.Lett. {\bf 74}, 2626(1995);\\
                D0 Collaboration,  Phys.Rev.Lett. {\bf74}, 2632(1995).
\item[{\rm [2]}] {\it For a review see}: W.Bernreuther et al., 
                 in {\it $e^+e^-$ Collisions at 500 GeV:
                 The Physics Potential}, P.Zerwas (Ed.), DESY 92-123A(1992),
		 Vol.I, p.255;\\
                 R.M.Barnett, R.Cruz, J.F.Gunion and B.Hubbard, 
                 Phys.Rev.D47(1993)1048;\\
		 K.Hidaka, Y.Kizukuri and T.Kon, Phys.Lett.B278(1992)155;\\
	         H.Baer, M.Drees, R.Godbole, J.F.Gunion and X.Tata, 
		 Phy.Rev.D44(1991)725;\\
		 M.Dree and D.P.Roy, Phys.Lett.B269(1991)155;\\
                 R.M.Godbole and D.P.Roy, Phys.Rev.D43(1991)3640;\\
                 V.Barger and R.J.N.Phillips, Phys.Rev.D44(1990)884;\\
                 H.Baer and X.Tata, Phys.Lett.B167(1986)241;\\
                 F.Borzumati and N.Polonsky, Report TMU-T31-87/95, 
                 hep-ph/9602433.
\item[{\rm [3]}] L.Rolandi (ALEPH), H.Dijkstra (DELPHI), D.Strickland (L3),
                 G.Wilson (OPAL), Joint Seminar on the First Results of LEP1.5,
                 CERN, December 12, 1995;\\
                 ALEPH Coll., CERN-PPE/96-10, Jan. 1996;\\
                 L3 Coll., H.Nowak and A.Sopczak, L3 Note 1887, Jan. 1996;\\
                 Opal Coll., S.Asai and S.Komamiya, OPAL Physics Note PN-205,
                 Feb.1996.
\item[{\rm [4]}] The D0 Collaboration, FERMILAB-Conf-95/186-E, {\it proc. of
                 the 10th Topical Workshop on Proton-Antiproton Collider 
                 Physics}, FNAL(1995).
\item[{\rm [5]}] J.Jezabek and J.H.Kuhn, Nucl.Phys.B314(1989)1;\\
		 C.S.Li, R.J.Oakes and T.C.Yuan, phys.Rev.D43(1991)3759;\\
                 G.Eilam, R.R.Mendel, R.Migneron, A.Soni, 
                    Phys.Rev.Lett.66(1991)3105;\\
                 C.P.Yuan and T.C.Yuan, Phys.Rev.D44(1991)3603;\\
                 J.Liu and Y.P.Yao, Int.J.Mod.Phys.A6(1991)4925;\\
                 C. S. Li, Jin Min Yang and B. Q. Hu, Phys Rev. D48(1993)
                 5425;\\                 
                 J.M.Yang and C.S.Li, Phys.Lett.B320(1994)117.
\item[{\rm [6]}] C.S.Li,  T.C.Yuan, Phys.Rev.D42(1990)3088; 
                   {\it} idid. 47(1993)2156(E);\\
                 C.S.Li, Y.S.Wei and J.M.Yang, Phys.Lett.B285(1992)137;\\
                 C.S.Li, B.Q.Hu and J.M.Yang, Phys.ReV.D47(1993)2865;\\
                 J.Liu and Y.P.Yao, Phys.ReV.D46(1992)5196;\\
                 A.Czarnecki and S.Davison, Phys.ReV.D48(1993)4183;\\
	         J.Guasch, R.A.Jimenez and J.Sola, Phys.Lett.B360(1995)47.
\item[{\rm[7]}]  H. E. Haber and G. L. Kane,  Phys. Rep. 117(1985)75;\\
                 J. F. Gunion and H. E. Haber, Nucl. Phys. B272(1986)1.
\item[{\rm[8]}]  J.Ellis and S.Rudaz, Phys.Lett.B128, 248 (1983)\\
                 A.Bouquet, J.Kaplan and C.Savoy, Nucl.Phys.B262, 299 (1985).
\item[{\rm[9]}]  W. Siegel, Phys.Lett.B84(1979)193;\\
                 D.M.Capper, D.R.T. Jones, P.van Nieuwenhuizen, Nucl. Phys.
                 B167(1980)479;\\
                 S. Martin and M.Vaugh, Phys. Lett. B318(1993)331.
\item[{\rm[10]}] A. Sirlin, Phys. Rev. D22(1980)971;\\
                 W.J. Marciano and A. Sirlin, {\it ibid.} 22(1980)2695; 
                                                        31,(1985)213(E);\\
		A. Sirlin and W.J. Marciano, Nucl.Phys.B189(1981)442;\\                
                K.I.Aoki et al., Prog.Theor.Phys.Suppl. 73(1982)1.
\item[{\rm[11]}] H.Eberl, A.Bartl and W. Majerotto, UWThPh-1996-6, hep-ph/9603206.
\item[{\rm[12]}] G. Passarino and M. Veltman, Nucl. Phys. B160(1979)151.
\item[{\rm[13]}] T.Kinosita, J.Math.Phys.3(1962)650;\\
                 T.D.Lee and M.Nauenberg, Phys.Rev.133, B1549(1964).
\item[{\rm[14]}] A.Denner and T.Sack, Z.Phys.C46(1990)653;\\
                 A.Denner, Fortschr.Phys.41(1993)4.                                         
\item[{\rm[15]}]  A. Sirlin, Phys. Rev. D{\bf 22}, 971 (1980);\\
            W. J. Marciano and A. Sirlin, Phys. Rev. D{\bf 22}, 2695 (1980);
            (E) D{\bf 31}, 213 (1985);\\
            A. Sirlin and W. J. Marciano, Nucl. Phys. {\bf B189}, 442 (1981);\\
            M. B\"ohm, W. Hollik and H. Spiesberger, Fortschr. Phys. {\bf 34},
            687 (1986).
\item[{\rm[16]}] W. J. Marciano and Z. Parsa, Annu. Rev. Nucl. Sci.36(1986)171.
  
\end{description}
\end{itemize}
\vfil
\eject

\begin{center} {\bf Figure Captions} \end{center}

Fig.1 Feynman diagrams for the  tree-level process $t\rightarrow \tilde t_1
\tilde \chi^0_j$ and the QCD corrections.

Fig.2 Feynman diagrams for the SUSY-QCD corrections.

Fig.3 The relative correction $\delta\Gamma/\Gamma_0$  to the decay rate 
as a function of the  
lighter stop mass assuming $m_{\tilde g}=500$GeV and $\tan\beta=11$.
The solid and dotted curves correspond to  $A_t+\mu\cot\beta=0$ (no mixing)
and $A_t+\mu\cot\beta=100$GeV (mixing), respectively.

Fig.4 The relative correction $\delta\Gamma/\Gamma_0$  to the decay rate 
as a function of the  
gluino mass assuming $m_{\tilde t_1}=50$GeV and $\tan\beta=11$.
The solid and dotted curves correspond to  $A_t+\mu\cot\beta=0$ (no mixing)
and $A_t+\mu\cot\beta=100$GeV (mixing), respectively.

Fig.5  The relative correction $\delta\Gamma/\Gamma_0$  to the decay rate 
as a function of 
$\tan\beta$ assuming  $m_{\tilde g}=500$GeV, $m_{\tilde t_1}=50$ GeV
and $A_t+\mu\cot\beta=100$GeV.

\end{document}